\documentclass{article}
\parskip=\medskipamount

\title{Conversion of second class constraints by deformation of Lagrangian local symmetries}
\author{A.A. Deriglazov\thanks{alexei.deriglazov@ufjf.edu.br ~ On leave of
absence from Dept. Math. Phys., Tomsk Polytechnical University,
Tomsk, Russia.}, ~ and Z. Kuznetsova}
\date{Dept. de Matematica, ICE, Universidade Federal de Juiz de Fora,\\
MG, Brazil; \\
and \\
LAFEX - CBPF/MCT, Rio de Janeiro, RJ, Brazil.}
\begin{document}
\maketitle
\large
\begin{abstract}
For a theory with first and second class constraints, we propose a procedure for conversion of second 
class constraints based on deformation the structure of local symmetries of the Lagrangian formulation. 
It does not require extension or reduction of configuration space of the theory. 
We give examples in which the initial formulation implies a non linear realization of some   
global symmetries, therefore is not convenient. The conversion reveals hidden symmetry 
presented in the theory. The extra gauge freedom of conversed version is used to search for a parameterization 
which linearizes the equations of motion. We apply the above procedure to membrane theory 
(in the formulation with world-volume metric). In the resulting version, all the metric components are gauge 
degrees of freedom. The above procedure works also in a theory with only second 
class constraints presented. As an examples, we discuss arbitrary dynamical system of classical 
mechanics subject to kinematic constraints, $O(N)$-invariant nonlinear sigma-model, and the theory 
of massive vector field with Maxwell-Proca Lagrangian.  
\end{abstract}

\noindent

\section{Introduction and outlook}
The conventional method for covariant quantization of a theory with second class constraints is to go over to 
an equivalent formulation where second class constraints are replaced by the first class ones in one or 
another way. One possibility is to work in extended  
phase space, where an additional variables can be used for conversion [1]. Another possibility is to search 
for special deformation of a theory in original  
phase space, which allows one either to discard half of the constraints ("gauge unfixing method") [2], 
or to solve the constraints in terms of a redundant number of variables [3]. Then the gauge theory 
quantization methods can be applied to the resulting formulation. 

The above mentioned conversion schemes have been developed in Hamiltonian framework. In this work we 
propose a conversion scheme based on a Lagrangian formulation. It does not imply a change (extension or reduction) 
of the number of configuration space variables. Roughly speaking, in Lagrangian theory (with first and second class 
constraints presented in the Hamiltonian formulation) we search for  
parameterization of configuration space which results in special deformation of original local symmetries 
and, in turn, implies conversion of second class constraints\footnote{Green-Schwarz superstring action can 
be interesting in this respect. While for $IIB$ case fermionic constraints can be covariantly separated into 
irreducible first and second class subsets [17], type $IIA$ and $N=1$ cases remain unsolved problem up to date.}.

The work is organized as follows. In the rest of this section we describe our procedure in some details.
In sections 2 and 3 we discuss two specific mechanical models where conversion of second class constraints 
allows one to make manifest hidden global symmetries of the theory. We point also that 
extra gauge freedom presented in the converted version can be useful for the linearization of equations of motion. 
In section 4 we convert second class constraints of membrane theory, in the formulation which involves 
world-volume metric. In resulting version all the metric components turn out to be gauge degrees of freedom. 
Further we explain how the 
conversion procedure can be applied in a theory with second class constraints only. Here our scheme implies 
an extension of original space by pure gauge degrees of freedom.
For the theory of massive 
vector field with Maxwell-Proca Lagrangian it simply reduces to introduction of Stuckelberg field (section 6).
Arbitrary dynamical system subject to kinematic constraints is considered in section 5, as particular 
examples we discuss a particle on a sphere and $O(N)$-invariant non linear sigma-model. 

Here we describe schematically our procedure of conversion. Let $L(q^A, \dot q^A)$ be 
Lagrangian of a theory with first and second class constraints presented in Hamiltonian formulation. In 
Lagrangian formulation, the first class constraints manifest themselves in invariance of action under some 
local symmetry transformations [4, 5, 6]. Let 
\begin{eqnarray}\label{01}
\delta q^A={\stackrel{(k)}{\epsilon}}R^A(q, \dot q)+ \ldots,
\end{eqnarray}
be infinitesimal form of one of the symmetries, with local parameter $\epsilon(\tau)$ and gauge generator  
$R^A$. In Eq.(\ref{01}) ${\stackrel{(k)}{\epsilon}}\equiv\frac{\partial^k\epsilon}{\partial\tau^k}$, and 
the dots stand for 
all terms with less then $k$-derivatives acting on a parameter. A local symmetry with at most $k$ 
derivatives acting on a parameter is called ${\stackrel{(k)}{\epsilon}}$-{\it symmetry} below. 
${\stackrel{(k)}{\epsilon}}$-symmetry generally implies [7] the appearance of some 
constraint on the $(k+1)$-stage of the Dirac procedure (clearly, it means that there is a  
chain [8] composed of primary, secondary, $\ldots$, $(k+1)$-stage constraints). This observation will be  
crucial for our present discussion. Now we describe how the symmetry (\ref{01}) can be used to convert 
some pair of second class constraints into a first class constraint.

Let us divide coordinates $q^A$ in two groups: $q^A=(q^i, ~ q^\alpha)$. We change parameterization of the  
configuration space: $q^A\longrightarrow\tilde q^A$ according to the transformation which involves 
derivatives of $q^\alpha$
\begin{eqnarray}\label{02}
q^i=q^i(\tilde q^A, ~ \dot{\tilde q}^\alpha), \qquad q^\alpha=q^\alpha(\tilde q^\beta).
\end{eqnarray}
We suppose that the transformation is "invertible"  
\begin{eqnarray}\label{03}
\det\frac{\partial q^i}{\partial\tilde q^j}\ne 0, \qquad 
\det\frac{\partial q^\alpha}{\partial \tilde q^\beta}\ne 0,
\end{eqnarray}
which implies that $\tilde q^A$ can be determined from (\ref{02}):
$\tilde q^i=\tilde q^i(q^A, ~ \dot q^\alpha)$, $\tilde q^\alpha=\tilde q^\alpha(q^\beta)$. 
Owing to the conditions (\ref{03}), our theory can be equally analyzed in terms of the Lagrangian 
$\tilde L\equiv L(q(\tilde q), \dot q(\tilde q))$. We further suppose that the transformation (\ref{02}) 
has been chosen 
in such a way that $\tilde L$ does not involve higher derivatives, modulo to total derivative term (we show below 
that it is possible in singular theory) 
\begin{eqnarray}\label{04}
\tilde L(\tilde q, \dot{\tilde q}, \ddot{\tilde q})=\tilde L'(\tilde q, \dot{\tilde q})+
\frac{dF(\tilde q, \dot{\tilde q})}{d\tau}.
\end{eqnarray}
Let us see what one can say about the structure of Hamiltonian constraints of our theory in the new 
parameterization $\tilde L$, in comparison with $L$. One should note that the local 
symmetry for the set $\tilde q$ is generally of ${\stackrel{(k+1)}{\epsilon}}$-type:  
$\delta \tilde q^i={\stackrel{(k+1)}{\epsilon}}\frac{\partial\tilde q^i}{\partial\dot q^\alpha}
\tilde R^\alpha(\tilde q^A, ~ \dot{\tilde q}^A, ~ \ddot{\tilde q}^\alpha)+\ldots$. Since the order of the symmetry has 
been raised by one unit, on $(k+2)$-stage of the Dirac procedure an extra constraint appears. 
On other hand, the physical sector of $\tilde L$ is the same as for $L$. If order of other symmetries (if any) was 
not lowered, the only possibility\footnote{Here the condition (\ref{04}) is important. A deformed 
theory with higher derivatives, being equivalent to the initial one, has more degrees of freedom than the number of 
variables $q^A$ [5]. So 
the extra constraints would be responsible for ruling out of these hidden degrees of freedom. 
Our condition (\ref{04}) forbids the appearance of the hidden degrees of freedom.} is 
that extra $(k+2)$-stage constraint is of first class, and it replaces some pair of second 
class constraints of initial formulation. In resume, an appropriate parameterization (\ref{02}), (\ref{03}), 
(\ref{04}) of the configuration space implies a deformation of local symmetries which, in turn, can result in 
conversion of second class constraints. Clearly, Eqs.({\ref{03}), (\ref{04}) represent only necessary 
conditions for the conversion. 

Note that one can consider more general transformations:    
$q^i=q^i(\tilde q^A, ~ \dot{\tilde q}^\alpha, ~ 
\ddot{\tilde q}^\alpha, \ldots, ~ {\stackrel{(s)}{\tilde q^\alpha}}), ~  
q^\alpha=q^\alpha(\tilde q^\beta)$ which involve higher derivatives of $\tilde q^\alpha$. It generally increases the  
order of symmetry by $s$ units, and $2s$ second class constraints can be converted. Example of such a kind is 
presented in the subsection 2.2. 

To illustrate the prescription, let us analyze the following dynamically trivial model defined on 
configuration space 
$x(\tau), ~ y(\tau), ~ z(\tau)$, with the Lagrangian action being 
\begin{eqnarray}\label{1}
S=\int d\tau\left(\frac{1}{2}(\dot x-y)^2+\frac{1}{2}z^2\right). 
\end{eqnarray}
It is invariant under finite local symmetry with the parameter $\alpha(\tau)$ 
\begin{eqnarray}\label{2}
\delta x=\alpha, \qquad \delta y=\dot\alpha, \qquad \delta z=0.
\end{eqnarray}
In terms of the variable set $x, ~y, ~ z$, the action (\ref{1}) has $\dot\alpha$-symmetry.
Passing to the Hamiltonian formulation, one obtains the following chains of constraints:
\begin{eqnarray}\label{3}
\quad\qquad \mbox{primary}\quad\ \qquad \mbox{secondary}\qquad \qquad \quad\cr
\mbox{first class chain }\qquad \qquad   p_y=0, \quad   \qquad  p_x=0, \qquad \qquad \qquad
\end{eqnarray}
\begin{eqnarray}\label{4}
\mbox{second class chain }~\qquad \quad    p_{ z}=0, \qquad \quad   z=0. ~ \qquad \qquad \qquad
\end{eqnarray}
To convert the second class constraints (\ref{4}), we make the transformation $z=\tilde z +\dot y$. In 
terms of $x, ~ y, ~ \tilde z$ variables, the action acquires the form
\begin{eqnarray}\label{5}
S=\int d\tau\left(\frac{1}{2}(\dot x-y)^2+\frac{1}{2}(\tilde z+\dot y)^2\right), 
\end{eqnarray}
and has irreducible $\ddot\alpha$-symmetry
\begin{eqnarray}\label{6}
\delta x=\alpha, \qquad \delta y=\dot\alpha, \qquad \delta\tilde z=-\ddot\alpha.
\end{eqnarray}
The necessary conditions (\ref{03}), (\ref{04}) are satisfied, and since order of symmetry has been raised, 
one expects that second class chain is replaced now by some tertiary first class constraint. Actually, 
the action (\ref{5}) implies the following first class chain 
\begin{eqnarray}\label{7}
p_{\tilde z}=0, \qquad p_y=0, \qquad p_x=0.
\end{eqnarray}
In the gauge $\tilde z=0$, the theories (\ref{1}) and (\ref{5}) have the same dynamics and thus   
are equivalent.
This example demonstrate also that our procedure is different from the conversion scheme of the work [3] 
based on a redundant 
parametrization\footnote{Our new variables $\tilde z, ~ p_{\tilde z}$ do not solve the second class 
constraints.}. In resume, second class constraints have been 
converted without changing (extension or reduction) of number of variables of the theory.

The condition (\ref{04}) can be easily satisfied if some variable enters into the action without derivative. 
In this respect, let us point out that for a singular theory $L(q, \dot q)$, there exists an equivalent formulation 
$L'(q', \dot q')$ with the desired property. Actually, starting from the singular $L$: 
$rank \frac{\partial^2 L}{\partial q^A\partial q^B}=[\alpha]<[A]$, 
one can construct the Hamiltonian $H=H_0(q^A, p_j)+v^\alpha\Phi_\alpha$, where 
$\Phi_\alpha(q^A, p_B)=p_\alpha-f_\alpha(q^A, p_j)$ are primary constraints, and the variables $q^A$ have been 
divided in two groups according to the rank condition: $q^A=(q^i, ~ q^\alpha)$, 
$\det\frac{\partial^2 L}{\partial q^i\partial q^j}\ne 0$. Here $H_0, ~ f_\alpha$ do not depend on $p_\alpha$ [9].
We further separate a phase space pair which corresponds to some fixed $\alpha$, for example $\alpha=1$: 
$\alpha=1, \alpha '$, 
$(q^A, p_A)=(q^1, p_1, z)$. According to [5] (see p. 256), there exists a canonical 
transformation $(q^1, p_1, z)\rightarrow (q'^{1}, p'_1, z')$, such that the Hamiltonian acquires the form 
$H'=H'_0(q'^{1}, z')+v^1p'_1+v^{\alpha'}\Phi_{\alpha'}(q'^{1}, z')$. One can 
restore [10] the Lagrangian $L'(q', \dot q')$ which reproduce $H'$ in the 
Hamiltonian formalism. By construction, $L^{'}$ does not depend on $\dot q'^1$.

\section{Conversion in a theory with hidden $SO(1, ~ 4)$ global symmetry}
Let's consider a theory with configuration space variables $x^\mu, ~ e, ~ g$ 
(where $\mu=0, 1, 2, 3, ~ \eta_{\mu \nu}=(-, +, +, +)$), and action  
\begin{eqnarray}\label{8}
S=\int d\tau\left(\frac{1}{2e}(\dot x^\mu-gx^\mu)^2+
\frac{1}{2e^2}g^2-ag\right), \qquad a=const. 
\end{eqnarray}
The model has a manifest $SO(1, 3)$ global symmetry. The only local symmetry is the reparametrization 
invariance, with form transformations being $\dot\alpha$-symmetry
\begin{eqnarray}\label{9}
\delta\tau=0, \qquad \delta x^\mu=-\alpha\dot x^\mu, \qquad 
\delta e=-(\alpha e)^{.}, \qquad \delta g=-(\alpha g)^{.}
\end{eqnarray}
The model turns out to be interesting in the context of doubly special relativity [11].
Passing to the Hamiltonian formalism one obtains the Hamiltonian ($v_i$ denote Lagrangian multipliers for the 
corresponding primary constraints)
\begin{eqnarray}\label{10}
H=\frac{e}{2} p^2+g(xp)-
\frac{g^2}{2e^2}+ag+v_{e}p_{e}+v_{g}p_{g},
\end{eqnarray}
as well as the constraints (the initial constraints have been reorganized with the aim to separate the first 
class ones)
\begin{eqnarray}\label{11}
p_{e}+(xp+a)p_{g}=0, \qquad p^2+(xp+a)^2+2ep^2p_{g}=0;
\end{eqnarray}
\begin{eqnarray}\label{12}
p_g=0, \qquad \qquad g-e(xp+a)=0.
\end{eqnarray}
The first (second) line represents first (second) class constraints. The equations of motion 
of the $(e, x)$-sector can be written as follow 
\begin{eqnarray}\label{13}
\dot e=v_{e}, ~  \quad 
\dot p_{e}=0, \qquad \qquad \qquad \qquad \qquad \qquad \qquad \cr
\dot x^\mu=e(p^\mu+(xp+a)x^\mu), \qquad 
\dot p^\mu=-e(xp+a)p^\mu.
\end{eqnarray}
In terms of variables
\begin{eqnarray}\label{14}
\mathcal X^\mu=\frac{ax^\mu}{xp+a}, \qquad 
\mathcal P^\mu=\frac{ap^\mu}{xp+a}, 
\end{eqnarray}
they acquire a form similar to those of free relativistic particle, namely 
\begin{eqnarray}\label{15}
\dot{\mathcal X}^\mu=e\mathcal P^\mu, \qquad 
\dot{\mathcal P}^\mu=0, \qquad \mathcal P^2=-a^2.
\end{eqnarray}
The presence of the conserved current $\dot{\mathcal P}^\mu=0$ indicates on hidden global symmetry related with  
the homogeneity of the configuration space. As it will be demonstrated, a conversion reveals the 
symmetry and allows one to find manifestly invariant formulation of the theory.

To convert a pair of second class constraints (\ref{12}) one needs to raise order of symmetry (\ref{9}) by unit. 
From Eq.(\ref{9}) one notes that it can be achieved by shifting some variable on $\dot e$. Since the 
variable $g$ enters into the action without derivative, a shift of the type $g=\tilde g+\dot e$ does not lead to 
higher derivative terms in the action and thus realizes the conversion. It is convienient to accompany the shift 
by an appropriate change of variables. Namely, let us make the invertible transformation 
$(x^\mu, e, g)\longrightarrow (\tilde x^A=(\tilde x^\mu, \tilde x^4), ~ \tilde g)$, where
\begin{eqnarray}\label{16}
\tilde x^\mu=e^{-\frac{1}{2}}x^\mu, \qquad 
\tilde x^4=e^{-\frac{1}{2}}, \qquad 
\tilde g=g-\frac{\dot e}{2e}.
\end{eqnarray}
In terms of these variables the action (\ref{8}) acquires the form 
\begin{eqnarray}\label{17}
\tilde S=\int d\tau\left(\frac{1}{2}(\dot{\tilde x}^A-\tilde g\tilde x^A)^2-a\tilde g\right), \qquad 
\eta_{AB}=(-,+,+,+,+), 
\end{eqnarray}
where the einbein $e$ was combined with $\tilde x^\mu$ to form a $5$-vector. The resulting action has a 
manifest $SO(1, 4)$ global symmetry. The conserved current $\mathcal P^\mu$ then corresponds 
to the symmetry under rotations in $(\tilde x^\mu, \tilde x^4)$-planes. THE Local symmetry of the action (\ref{17}) 
can be obtained from Eqs. (\ref{9}), (\ref{16}), and is of $\ddot\alpha$-type
\begin{eqnarray}\label{18}
\delta\tau=0, \qquad \delta\tilde x^A=\frac{1}{2}\dot\alpha\tilde x^A-\alpha\dot{\tilde x}^A, \qquad
\delta \tilde g=\frac12\ddot\alpha-\dot\alpha{\tilde g}-\alpha\dot{\tilde g}.
\end{eqnarray}
Passing to the Hamiltonian formulation one obtains the Hamiltonian
\begin{eqnarray}\label{19}
H=\frac12\tilde p^2+\tilde g\tilde x^A\tilde p_A+a\tilde g+v_{\tilde g}p_{\tilde g},
\end{eqnarray}
and the constraints
\begin{eqnarray}\label{20}
\tilde p_{\tilde g}=0, \qquad \tilde x^A\tilde p_A+a=0, \qquad \tilde p^A\tilde p_A=0,
\end{eqnarray}
all of them being the first class. Thus $\tilde S$ represents the converted version of the action(\ref{8}). 
Let us write equations of motion for $x^A$-sector
\begin{eqnarray}\label{21}
\dot{\tilde x}^A=\tilde p^A+\tilde g\tilde x^A, \qquad \dot{\tilde p}^A=-\tilde g\tilde p^A.
\end{eqnarray}
In the gauge $\tilde g=\tilde x^\mu \tilde p_\mu+a, ~ \tilde p_4=\tilde x^\mu \tilde p_\mu+a$ for 
the theory (\ref{17}) one reproduces the 
initial dynamics (\ref{13}) (taken in the gauge $e=1$).  Going over to the gauge $\tilde g=0, ~ \tilde p_4=a$, one 
obtains the free equations (\ref{15}). 
Hence the extra gauge freedom, resulting from the conversion of second class constraints, can be used for search for 
parametrization which linearises equations of motion.

\section{Conversion in a theory with hidden conformal symmetry}

Here we discuss a conversion of a chain with four second class constraints presented. Let us consider an action with 
manifest $SO(1, 4)$ global symmetry 
\begin{eqnarray}\label{22}
S=\int d\tau\left(\frac{1}{2e}(\dot x^A)^2-
\frac{e}{2}m^2+g\left((x^A)^2-a^2\right)\right),
\end{eqnarray}
where $A, B=0,1,2,3,4, ~ \eta_{AB}=(-,+,+,+,+), ~ m, a=const, a\ne 0$. It is a reparametrization invariant, with the 
form transformations being $\dot\alpha$-symmetry, see (\ref{9}). In the Hamiltonian formulation one finds the 
following system of constraints 
\begin{eqnarray}\label{23}
p_e+\frac{m^2}{2a^2}p_{g}=0, \qquad (p^A)^2+m^2-
\frac{m^2}{a^2}((x^A)^2-a^2)=0;
\end{eqnarray}
\begin{eqnarray}\label{24}
p_g=0, \qquad (x^A)^2-a^2=0, \qquad x^Ap_A=0, \qquad 
g-\frac{m^2}{2a^2}e=0.
\end{eqnarray}
The first (second) line represents first (second) class constraints. The chain of four second class constraints  
can be converted by raising of order of the local symmetry by two units. To this end, let us make invertible 
transformation $(x^A, e, g)\longrightarrow (\tilde x^M=(\tilde x^A, \tilde x^5), ~ \tilde g)$, where 
\begin{eqnarray}\label{25}
\tilde x^A=e^{-\frac{1}{2}} x^A, \qquad 
\tilde x^5=ae^{-\frac{1}{2}}, \qquad 
\tilde g=eg+\frac{3\dot e^2}{8e^2}-\frac{\ddot e}{4e}.
\end{eqnarray}
For this set of variables, the action (\ref{22}) acquires the form (note that there are no of higher derivative 
terms) 
\begin{eqnarray}\label{26}
\tilde S=\int d\tau\left(\frac{1}{2}(\dot{\tilde x}^M)^2+\tilde g(\tilde x^M)^2-
\frac12 a^2m^2(\tilde x^5)^{-2}\right), \cr 
\eta_{MN}=(-,+,+,+,+,-). \qquad \qquad  
\end{eqnarray}
Local symmetry of (\ref{26}) can be obtained from Eqs. (\ref{9}), (\ref{25}), and is of 
$\stackrel{(3)}{\alpha}$-type
\begin{eqnarray}\label{27}
\delta\tau=0, \quad \delta \tilde x^M=\frac{1}{2}\dot\alpha \tilde x^M-\alpha\dot{\tilde x}^M, \quad
\delta \tilde g=\frac14\stackrel{(3)}{\alpha}-2\dot\alpha \tilde g-\alpha\dot{\tilde g}.
\end{eqnarray}
In the Hamiltonian formulation one obtains the constraints
\begin{eqnarray}\label{28}
\tilde p_{\tilde g}=0, \quad (\tilde x^M)^2=0, \quad \tilde x^M\tilde p_M=0, \quad 
(\tilde p_M)^2+c^2m^2(\tilde x^5)^{-2}=0, 
\end{eqnarray}
all of them being the first class. Thus the transformation (\ref{25}) turn out $\dot\alpha$-symmetry of the 
initial action into $\stackrel{(3)}{\alpha}$-symmetry, which results in replacement of four second class 
constraints (\ref{24}) by a pair of first class ones.

For completeness, let us compare equations of motion for the action $S$
\begin{eqnarray}\label{29}
\dot e=v_e, \qquad \qquad \qquad \dot p_e=0; \quad \cr 
\dot x^A=ep^A, \qquad \dot p^A=m^2a^{-2}ex^A,
\end{eqnarray}
with the corresponding equations for the action $\tilde S$
\begin{eqnarray}\label{30}
\dot{\tilde x}^5=\tilde p^5, \qquad \dot{\tilde p}^5=2\tilde g\tilde x^5-a^2m^2(\tilde x^5)^{-3}; \cr
\dot{\tilde x}^A=\tilde p^A, \qquad \dot{\tilde p}^A=2\tilde g\tilde x^A. \qquad \qquad ~ \quad
\end{eqnarray}
In the gauge $e=1$ for the first theory and $\tilde x^5=a, ~ \tilde p^5=0, ~ \tilde g=\frac{m^2}{2a^2}$ 
for the second theory 
the equations (as well as the remaining constraints) coincide. The constraints $(\tilde x^M)^2=0, ~ 
\tilde x^M\tilde p_M=0$ can also 
be linearized, see [12].

The action (\ref{22}) with $m=0$ implies conservation of $p^A$: $\dot p^A=0$, the latter equation 
appears as one of equations of motion. It indicates on hidden global symmetry responsible for the current. 
The conversion of second class constraints made by transition to the action (\ref{26}) reveals the symmetry: 
the action (\ref{26}) with $m=0$ is $SO(2, 4)$-invariant. The current $p^A$ corresponds to 
rotations in $(\tilde x^A, \tilde x^5)$-planes.

\section{Conversion of second class constraints in the membrane action}
Here we consider a membrane in terms of variables $x^\mu(\sigma^i), ~ g^{ij}(\sigma^i)$, 
where $\sigma^i, ~ i=0, 1, 2$ are coordinates parametrizing world-volume, $x^\mu, ~ \mu=0, 1, 2, \ldots, D-1$ 
gives embedding of the world-volume in a Minkowski space-time, $g^{ij}$ represent metric on the world-volume. 
The membrane action [13]
\begin{eqnarray}\label{31}
S=\frac{T}{2}\int d^3\sigma(-det g^{ij})^{-\frac{1}{2}}(-g^{ij}\partial_i x^\mu\partial_j x^\mu+1),
\end{eqnarray}
is invariant under reparametrizations on the world-volume, where $x^\mu$ are scalar functions and $g^{ij}$ 
is a second rank tensor. The corresponding infinitesimal transformations of the form are 
\begin{eqnarray}\label{32}
\delta\sigma^i=0, \quad \delta x^\mu=-\xi^i\partial_i x^\mu, \qquad \qquad \qquad \cr
\delta g^{ij}=g^{ik}\partial_k\xi^j+g^{jk}\partial_k\xi^i-\xi^k\partial_kg^{ij}=
g^{i0}\dot\xi^j+g^{j0}\dot\xi^i+\ldots ,
\end{eqnarray}
where in the second line we have omitted those terms which do not involve time derivative of parameters.
Owing to $\dot\xi$-symmetry (\ref{32}), six first class constraints appear in the Hamiltonian formulation. 
Besides (note that the metric obeys algebraic equations), more six constraints of second class are presented. 
We demonstrate below, how the second class constraints  can be converted into first class ones by deformation 
of local symmetry (\ref{32}).

We begin with making convenient parametrization of the world-volume metric. 
Namely, let's consider the following change of variables\footnote{they are related with conventional ADM variables
$g^{00}=-\tilde N^{-2}, ~ g^{0a}=\tilde N^{-2}\tilde N^a, ~ 
g^{ab}=\tilde\gamma^{ab}-\tilde N^{-2}\tilde N^a\tilde N^b$ as follows:
$\tilde N=(\det \gamma^{ab})^{\frac12}N^{-1}, ~ \tilde N^a=N^a N^{-1}, ~ 
\tilde\gamma^{ab}=(\det \gamma^{ab})^{-1}\gamma^{ab}$.}: $g^{ij}\longrightarrow (N, N^a, \gamma^{ab}),  
a, b=1, 2$, where 
\begin{eqnarray}\label{33}
g^{ij}=
\left(
\begin{array}{cc}
-(\det \gamma^{ab})^{-1}N^2&(\det \gamma^{ab})^{-1}N N^a\\
(\det \gamma^{ab})^{-1}N N^b&(det \gamma^{ab})^{-1}(\gamma^{ab}-N^aN^b)
\end{array}
\right).
\end{eqnarray}
It is invertible, with the inverse transformation being
\begin{eqnarray}\label{34}
N=g^{00}(-\det g^{ij})^{-\frac12}, \qquad N^a=g^{0a}(-\det g^{ij})^{-\frac12}, \cr
\gamma^{ab}=(\det g^{ij})^{-1}(g^{ab}g^{00}-g^{0a}g^{0b}).\quad
\end{eqnarray}
Now the action acquires a polynomial form for all variables\footnote{An additional transformation 
$N^{-1}N^a=\hat N^{a}, N\gamma^{ab}=\hat\gamma^{ab}$ implies polynomial form of action. But the hatted  
variables have more complicated transformation law.} except $N$
\begin{eqnarray}\label{35}
S=\frac{T}{2}\int d^3\sigma(N(\dot x^\mu-N^{-1}N^a\partial_ax^\mu)^2-
N^{-1}\gamma^{ab}\partial_a x^\mu\partial_b x^\mu+ \cr 
N^{-1}\det \gamma^{ab}) \qquad \qquad \qquad.
\end{eqnarray}
Moreover, the symmetry (\ref{32}) acquires more transparent form for the new variables, in 
particular, $\delta\gamma^{ab}$ do not involves time derivative of the parameters 
\begin{eqnarray}\label{36}
\delta N=N\dot\xi^0+\ldots , \qquad \delta N^a=N\dot\xi^a+\ldots ,\qquad \delta\gamma^{ab}=0+\ldots .
\end{eqnarray}
In the Hamiltonian formalism the action (\ref{35}) implies the constraints
\begin{eqnarray}\label{37}
p_N=0, \qquad \frac{p^2}{T^2}+\det(\partial_ax\partial_bx)=0, \cr 
p_{Na}=0, \qquad \qquad p\partial_ax=0, \qquad \quad
\end{eqnarray}
\begin{eqnarray}\label{371}
\pi_{ab}=0, \qquad (\det \gamma_{cd})^{-1}\gamma_{ab}=\partial_ax\partial_bx, 
\end{eqnarray}
where the last line represents six second class constraints. Here $\pi_{ab}$ are conjugated momenta for
$\gamma^{ab}$, and $\gamma_{ab}$ is inverse matrix for $\gamma^{ab}$. The last expression in (\ref{371}) 
is equivalent to $\gamma^{22}=\partial_1x\partial_1x, ~ \gamma^{12}=-\partial_1x\partial_2x, ~ 
\gamma^{11}=\partial_2x\partial_2x$. Eq. (\ref{36}) suggests that conversion can be performed by the 
following shift in Eq.(\ref{35})
\begin{eqnarray}\label{38}
\gamma^{ab}=
\left(
\begin{array}{cc}
h^{11}+\dot N^1&h^{12}+\dot N\\
h^{12}+\dot N&h^{22}+\dot N^2
\end{array}
\right).
\end{eqnarray}
In comparison with the initial action (\ref{35}), one has now kinetic terms for $N$-fields. So the only three 
primary constraints appear: $\pi_{ab}=0$, where $\pi_{ab}$ are conjugated momenta for $h^{ab}$. On the other 
hand, the modified action has three $\ddot\alpha$-symmetries, see Eqs. (\ref{36}), (\ref{38}). Thus one 
expects appearance of three tertiary first class constraints, the latter replace six second class 
constraints (\ref{371}) of the initial formulation. In some details, for the modified action
\begin{eqnarray}\label{39}
\tilde S=\frac{T}{2}\int d^3\sigma\left[N(\dot x^\mu-N^{-1}N^a\partial_ax^\mu)^2-\right. \qquad \cr
\left. N^{-1}((h^{aa}+\dot N^a)\partial_a x\partial_a x+2(h^{12}+
\dot N)\partial_1 x\partial_2 x-\right.\cr 
\left. (h^{11}+\dot N^1)(h^{22}+\dot N^2)-(h^{12}+\dot N)^2)\right], \qquad
\end{eqnarray}
one obtains the Hamiltonian
\begin{eqnarray}\label{40}
H=\int d^2\sigma(\frac{1}{2TN}p^2+\frac{N^a}{N}p\partial_ax-\frac{N}{2T}p_{N}^2+
\frac{2N}{T}p_{N1}p_{N2}- \cr
p_Nh^{12}-p_{Na}h^{aa}-p_N\partial_1x\partial_2x+p_{N2}\partial_1x\partial_1x+ \qquad \quad \cr
p_{N1}\partial_2x\partial_2x+\frac{T}{2N}det(\partial_ax\partial_bx)+v_{h}^{ab}\pi_{ab}), \qquad \qquad 
\end{eqnarray}
as well as the following three chains of first class constraints
\begin{eqnarray}\label{41}
\pi_{12}=0, \qquad \qquad p_N=0, \qquad \frac{p^2}{T^2}+det(\partial_ax\partial_bx)=0, \cr 
\pi_{aa}=0, \qquad \qquad p_{Na}=0, \qquad \qquad p\partial_ax=0. \qquad \quad ~
\end{eqnarray}
Thus all the metric components turn out to be gauge degrees of freedom in the theory (\ref{39}). Starting from 
the Hamiltonian (\ref{40}), one obtains the well known membrane equations of motion [14] in the gauge 
$N=1, ~ N^a=0, ~ (deth_{ab})^{-1}h_{ab}=\partial_ax\partial_bx$ (they can be linearized for half-rigid 
membrane [15]).

In resume, we have found a special representation (\ref{33}), (\ref{38}) for the membrane world-volume metric. 
The reparametrization invariance for the new variables turns out to be a symmetry of $\ddot\alpha$-type, which implies 
conversion of second class constraints presented in the initial action. In the modified action (\ref{39}), 
all the metric components are gauge degrees of freedom. It would be interesting to find manifestly 
$\ddot\alpha$-symmetry covariant formulation for the action (\ref{39}).

\section{Classical mechanics subject to kinematic constraints as a gauge theory}
Our conversion trick can be realized also in a theory with second class constraints only (i. e. in a 
theory without local symmetries presented in the initial formulation). To proceed with, one notes that arbitrary 
theory without local symmetry can be treated as a gauge theory on appropriately extended configuration 
space. Namely, given theory with the action $S(q^A)$ on configuration space $q^A$ can be equally considered as a 
theory on the space $q^A, a$, with local transformations defined by $q'^A=q^A, a'=a+\alpha$, where $a$ is one more 
configuration space variable. Since $a$ does not enter into the action, the latter is invariant under the 
local transformations\footnote{It is general situation: for an arbitrary locally invariant theory 
one can chose special 
variables such that the action does not depend on some of them [5].}.
The trivial gauge symmetry of the extended formulation can be further used for the conversion of second class 
constraints according to our procedure\footnote{There are other possibilities to create trivial local symmetries. 
For example, in a given Lagrangian action with one of variables being $q$, let us make the substitution 
$q=ab$, where $a, ~b$ represent new configuration space variables. 
The resulting action is equivalent to the initial one, an auxiliary character of one of new degrees of 
freedom is guaranteed by the trivial gauge symmetry: 
$a\rightarrow a^{'}=\alpha a, ~ b\rightarrow b^{'}=\alpha^{-1}b$.   
Another simple possibility is to write $q=a+b$, which implies the symmetry 
$a\rightarrow a^{'}=a+\alpha, ~ b\rightarrow b^{'}=b-\alpha$. The well known examples of such a kind 
transformation are einbein formulation in gravity theory: $g_{\mu\nu}=e^a_\mu e^a_\nu$ (which implies 
local Lorentz invariance), as well as duality transformations in some specific models [16].}.    

Let us see how it works on example of classical mechanics with kinematic constraints. Let 
$L_0(q^a, ~ \dot q^b)$ be Lagrangian of some system of classical mechanics in terms of generalized 
coordinates $q^a$. The Lagrangian is supposed to be nondegenerate
\begin{eqnarray}\label{42}
\det\frac{\partial^2L_0}{\partial\dot q^a\partial\dot q^b}\ne0. 
\end{eqnarray}
A motion restricted to lie on some 
hypersurface defined by nondegenerate system of equations $G_i(q^a)=0, ~ 
rank\frac{\partial G_i}{\partial q^a}=[i]<[a]$ can be described by the well known action with Lagrangian 
multipliers $\lambda^i(\tau)$ 
\begin{eqnarray}\label{43}
S=\int d\tau(L_0(q, \dot q)+\lambda^iG_i(q)).
\end{eqnarray}
Here the variables $\lambda^i(\tau)$ are considered on equal footing with original variables $q^a(\tau)$. 
Let us construct a 
Hamiltonian description of the system. Due to the rank condition (\ref{42}), equations for the momenta: 
$p_a=\frac{\partial L_0}{\partial\dot q^a}$ can be resolved in relation of $\dot q^a$. 
Let $\dot q^a=f^a(q, p)$ be solution:
\begin{eqnarray}\label{44}
\left.\frac{\partial L_0}{\partial\dot q^a}\right|_{\dot q=f(q, p)}\equiv p_a, \qquad 
\det\frac{\partial f^a}{\partial p_b}\ne 0. 
\end{eqnarray}
Conjugated momenta for $\lambda^i$ represent $[i]$ primary constraints of the theory: $p_{\lambda i}=0$. 
Then one obtains the Hamiltonian  
\begin{eqnarray}\label{45}
H=H'-\lambda^iG_i(q)+v_{\lambda}^ip_{\lambda i}, \qquad  
H'\equiv p_af^a-L(q, f).
\end{eqnarray}
Conservation in time of the primary constraints: $\dot p_{\lambda i}=\{p_{\lambda i}, H\}=0$ implies secondary 
constraints $G_i(q)=0$. In turn, conservation of $G$ gives tertiary constraints $F_i\equiv G_{ia}(q)f^a(q, p)=0$, 
where $G_{ia}\equiv\frac{\partial G_i}{\partial q^a}$ and Eq.(\ref{44}) was used. Poisson brackets of the 
constraints are  
\begin{eqnarray}\label{46}
\{G_ i, F_ j\}=G_{ia}\frac{\partial f^c}{\partial p_a}G_{jc}\equiv\triangle_{ij}.
\end{eqnarray}
Owing to Eq.(\ref{44}) and the condition $rank G_{ia}=[i]$ one concludes $\det\triangle_{ij}\ne 0$. 
The inverse matrix for $\triangle$ is denoted as $\tilde\triangle^{ij}$. Further, the condition 
$\dot F_i=0$ implies fourth stage constraints $\lambda^i-\tilde\triangle^{ij}\{F_j, H'\}=0$. At last, 
conservation in time 
of these constraints determines all the remaining velocities: 
$v_{\lambda}^i=\{\tilde\triangle^{ij}\{F_j, H'\}, H'-\lambda^kG_k\}$. 
Thus we have a theory with $4[i]$ second class constraints 
\begin{eqnarray}\label{47}
p_{\lambda i}=0, \quad G_i=0, \quad f^aG_{ia}=0, \quad \lambda^i-\tilde\triangle^{ij}\{F_j, H'\}=0.
\end{eqnarray}
The conversion can be carried out by making of the following transformation in the action (\ref{43})
\begin{eqnarray}\label{48}
\lambda^i=\tilde\lambda^i+\ddot e^i,
\end{eqnarray}
where auxiliary configuration space variable $e^i(\tau)$ has been introduced. The modified action
\begin{eqnarray}\label{49}
\tilde S=\int d\tau(L_0(q, \dot q)-\dot e^iG_{ia}\dot q^a+\tilde\lambda^iG_i(q)),
\end{eqnarray}
does not contain higher derivative terms and is invariant under local transformations 
$\tilde\lambda^i\rightarrow\tilde\lambda^{'i}=\tilde\lambda^i+\ddot\alpha^i, ~ 
e^i\rightarrow e^{'i}=e^i-\alpha^i$. Due to this $\ddot\alpha$-symmetry one expects an appearance of $3[i]$ 
first class constraints in the Hamiltonian formulation for the theory (\ref{49}). To confirm this, let us 
write defining equations for conjugated momenta
\begin{eqnarray}\label{50}
p_a\equiv\frac{\partial L}{\partial\dot q^a}=\frac{\partial L_0}{\partial\dot q^a}-
\dot e^iG_{ia}, \quad p_{ei}\equiv\frac{\partial L}{\partial\dot e^i}=-G_{ia}\dot q^a, 
\quad p_{\tilde\lambda i}=0.
\end{eqnarray}
The last equation represents $[i]$ primary constraints. Remaining equations can be resolved in 
relation of the velocities $\dot q^a, \dot e^i$, since the corresponding block of Hessian matrix is non 
degenerate. It can be easily seen in special coordinates chosen as 
follows. The initial coordinates $q^a$ can be reordered in such a way that rank minor of the matrix 
$\frac{\partial G_i}{\partial q^a}$ is placed on the right: $ q^a=(q^\alpha, ~ q^i), ~
\det\frac{\partial G_i}{\partial q^j}\ne 0$. Now, let us make invertible change of variables 
$q^a\rightarrow \tilde q^a$, where $\tilde q^{\alpha}=q^{\alpha}, ~ \tilde q^i=G_i(q^a)$. In this variables 
our Lagrangian is   
\begin{eqnarray}\label{51}
\tilde L=L_0(\tilde q, \dot{\tilde q})-\dot e^i\dot{\tilde q}^i+\tilde\lambda^i\tilde q^i.
\end{eqnarray}
From this expression one immediately finds the determinant of the Hessian matrix being 
$\det\frac{\partial^2\tilde L}{\partial^2(\tilde q, e)}=
\det\frac{\partial^2L_0}{\partial\dot{\tilde q}^\alpha\partial\dot{\tilde q}^\beta}$. It is nonzero since in 
classical mechanics the quadratic form 
$\frac{\partial^2L_0}{\partial\dot{\tilde q}^a\partial\dot{\tilde q}^b}$ is positive defined. 

Let us return to analysis of the action (\ref{49}). The corresponding Hamiltonian is 
\begin{eqnarray}\label{52}
H=p_a\dot q^a+p_{ei}\dot e^i-L_0(q, ~ \dot q)+\dot e^iG_{ia}q^a-\tilde\lambda^iG_i(q)+
v_{\tilde\lambda}^ip_{\tilde\lambda i},
\end{eqnarray}
where $\dot q^a, ~ \dot e^i$ are solutions of equations (\ref{50}). As before, secondary constraints 
turn out to be $G_i(q)=0$. Their conservation in time can be easily computed by using of Eq.(\ref{50}): 
$\dot G_i=\{G_i, ~ H\}=-p_{ei}$ which gives tertiary constraints $p_{ei}=0$. Then the complete constraint 
system is composed by $3[i]$ first class constraints
\begin{eqnarray}\label{53}
p_{\tilde\lambda i}=0, \qquad G_i=0, \qquad p_{e i}=0.
\end{eqnarray}
First class constraint $p_{e i}=0$ simply states that variables $e^i$ are pure gauge degrees of freedom, as 
it was expected. 
They can be removed from the formulation if one chooses the gauge $e^i=0$. The remaining $2[i]$ first class 
constraints in Eq.(\ref{53}) replace $4[i]$ second class constraints (\ref{47}) of the initial formulation.  

As a particular example, let us consider a motion of a {\bf particle on $2$-sphere of radius $c$}, with the 
action being 
\begin{eqnarray}\label{531}
S=\int d^3x\left(\frac12m(\dot x^i)^2+\lambda((x^i)^2-c^2)\right).
\end{eqnarray}
It implies the following chain of 4 second class constraints
\begin{eqnarray}\label{532}
p_{\lambda}=0, \qquad x^2-c^2=0, \qquad xp=0, \qquad p^2+2mc^2\lambda=0.
\end{eqnarray}
The conversion is achieved by the transformation $\lambda=\tilde\lambda+\frac12m\ddot e$, which generates the symmetry 
$\tilde\lambda\rightarrow\tilde\lambda^{'}=\tilde\lambda+\frac12m\ddot\alpha, ~ e\rightarrow e^{'}=e-\alpha$. 
The modified action  
\begin{eqnarray}\label{533}
\tilde S=\int d^3x\left(\frac12m\dot x^2-m\dot ex\dot x+\tilde\lambda(x^2-c^2)\right).
\end{eqnarray}
implies first class constraints only, namely
\begin{eqnarray}\label{534}
p_{\tilde\lambda}=0, \qquad x^2-c^2=0, \qquad p_e=0.
\end{eqnarray}
{\bf $O(N)$-invariant  nonlinear sigma-model} 
\begin{eqnarray}\label{535}
S=\int d^Dx\left(\frac12(\partial_\mu \phi^a)^2+\lambda((\phi^a)^2-1)\right),
\end{eqnarray}
represents example of field theory with similar structure of second class constraints. Hence the transformation 
$\lambda=\tilde\lambda+\partial_\mu\partial^\mu e$ gives formulation with first class constraints only
\begin{eqnarray}\label{536}
\tilde S=\int d^Dx\left(\frac12(\partial_\mu \phi^a)^2-2\partial_\mu e\partial^\mu\phi^a\phi^a+
\tilde\lambda((\phi^a)^2-1)\right).
\end{eqnarray}

\section{Conversion in Maxwell-Proca Lagrangian for massive vector field}
As one more example of the conversion in a theory with second class constraints only, we consider massive vector 
field $A^\mu(x^\nu)$ in Minkowski space (with the signature being $(-, +, +, +)$) . 
It is described by the following action:
\begin{eqnarray}\label{54}
S=\int d^4x(-\frac14F_{\mu\nu}F^{\mu\nu}+\frac12m^2A^\mu A_\mu), 
\quad F_{\mu\nu}\equiv\partial_\mu A_\nu-\partial_\nu A_\mu.
\end{eqnarray}
Passing to the Hamiltonian formulation one finds the Hamiltonian 
\begin{eqnarray}\label{55}
H=\int d^3x(\frac12p_i^2-p_i\partial_i A^0+\frac14F_{ij}^2-\frac12m^2A^\mu A_\mu+v_0p_0),
\end{eqnarray}
as well as primary and secondary constraints 
\begin{eqnarray}\label{56}
p_0=0, \qquad \partial_ip_i+m^2A^0=0.
\end{eqnarray}
The system is of second class, with the Poisson bracket algebra being
\begin{eqnarray}\label{57}
\{\partial_ip_i+m^2A^0, p_0\}=m^2\delta^3(x-y).
\end{eqnarray}
Conservation in time of the secondary constraint determines the velocity $v_0=-\partial_kA_k$. The equations of 
motion for propagating modes turn out to be 
\begin{eqnarray}\label{571}
\partial_0A^i=p^i-\partial_iA^0, \qquad 
\partial_0p_i=\partial_kF_{ki}+m^2A_i, 
\end{eqnarray}
while the modes $A^0, ~ p_0$ are determined by the algebraic equations 
(\ref{56}). In a converted version these modes turn into gauge degrees of freedom. For the case, a 
transformation which criates desirable $\dot\alpha$ - symmetry consist in introduction of Stuckelberg 
field $\phi(x^\mu)$ 
\begin{eqnarray}\label{58}
A_\mu=\tilde A_\mu-\partial_\mu\phi. 
\end{eqnarray}
According to our philosophy, one can think that, from the beginning, we have a theory on configuration space 
$A_\mu, \phi$, with the local symmetry being $A'^\mu=A^\mu$, $\phi '=\phi+\alpha$, and the action given 
by Eq. (\ref{54}). That is $\phi$ does not enter into the action. Then one introduces the new parametrization 
(\ref{58}) of the configuration space: $A_\mu, \phi \rightarrow \tilde A_\mu, \phi$. The modified action 
\begin{eqnarray}\label{59}
\tilde S=\int d^4x(-\frac14\tilde F_{\mu\nu}\tilde F^{\mu\nu}+\frac12m^2(\tilde A^\mu-
\partial^\mu\phi)(\tilde A_\mu-\partial_\mu\phi)), \cr 
\tilde F_{\mu\nu}=\partial_\mu \tilde A_\nu-\partial_\nu \tilde A_\mu, \qquad \qquad \qquad
\end{eqnarray}
is invariant under the local transformations
\begin{eqnarray}\label{60}
\phi\rightarrow \phi^{'}=\phi+\alpha, \qquad \tilde A_\mu\rightarrow \tilde A^{'}_\mu=\tilde A_\mu+\partial_\mu\alpha,
\end{eqnarray}
where $\tilde A_\mu$ transforms as electromagnetic field. Due to this $\dot\alpha$-symmetry, one expects 
appearance of two first class constraints in the modified formulation. Actually, primary constraint 
of the theory (\ref{59}) is the same as before: $\tilde p_0=0$. Then the Hamiltonian turns out to be  
\begin{eqnarray}\label{61}
H=\int d^3x\left(\frac12\tilde p_i^2-\tilde p_i\partial_i\tilde A^0+\frac14\tilde F_{ij}^2-\frac{1}{2m^2}p_{\phi}^2-
p_{\phi}\tilde A^0-\right. \cr
\left.\frac12m^2(\tilde A^i-\partial_i\phi)^2+v_0\tilde p_0\right), \qquad \qquad \quad
\end{eqnarray}
and implies secondary constraint $\partial_i\tilde p_i+p_{\phi}=0$. Complete constraint system  
\begin{eqnarray}\label{62}
\tilde p_0=0, \qquad \partial_i\tilde p_i+p_{\phi}=0,
\end{eqnarray}
is of first class. The last constraint in Eq. (\ref{62}) states that $\phi$ is an auxiliary degree of freedom. 
It can be removed by the gauge $\phi=0$. The first class constraint $\tilde p_0=0$ replaces two second 
class constraints (\ref{56}) 
of initial formulation, and states that $A^0$ is a gauge degree of freedom in the modified formulation (\ref{59}).
Equations of motion for propagating modes in the modified theory are slightly different
\begin{eqnarray}\label{63}
\partial_0\tilde A^i=\tilde p^i-\partial_i\tilde A^0, \qquad 
\partial_0\tilde p_i=\partial_k\tilde F_{ki}+m^2(\tilde A_i-\partial_i \phi).
\end{eqnarray}
Nevertheless, in the gauge $\phi=0$ they coincide with corresponding equations (\ref{571}) of initial formulation.

\section{Conclusion}
In this work we have proposed scheme for conversion of second class constraints which is mainly deal with the 
Lagrangian formulation of a theory. For the Lagrangian theory with first and second class constraints presented in 
the Hamiltonian formulation, it does not require neither increase nor decrease of number of initial variables. 
The scheme was developed on a base of the following observations. \par
\noindent
a) One can change parameterization of configuration space by making use of transformation which involves 
derivatives of variables, see Eq. (\ref{02}). The condition (\ref{03}) then guarantees that the theory can be 
equally analyzed in terms of the new variables. \par
\noindent
b) Such a kind transformation increase order of local symmetry: transformation law for the new variables 
involves more derivatives acting on the local symmetry parameters as compare with the initial 
formulation. \par 
\noindent
c) In turn, it generally implies appearance [7] of higher-stage first class constraints, the latter replace 
second class constraints of the original formulation.

While we have formulated only necessary conditions (\ref{03}), (\ref{04}) for our conversion scheme, its 
efficacy has been demonstrated on a number of examples. In particular, conversion of second class 
constraints for the membrane action (\ref{31}) as well as for an arbitrary dynamical system 
subject to kinematic constraints (\ref{43}) was not realized by the methods developed in [1-3].

\section{Acknowledgments}
A. A. D. would like to thank the Brazilian foundations CNPq (Conselho Nacional de Desenvolvimento 
Científico  e Tecnológico - Brasil) and FAPEMIG for financial support. Z. K. would like to thank the 
Brazilian foundation FAPEMIG.

\end{document}